\documentclass[preprint,showpacs,preprintnumbers,amsmath,amssymb,superscriptaddress]{revtex4}

\usepackage{graphicx,amsfonts}% Include figure files
\usepackage{epsfig}
\usepackage{dcolumn}% Align table columns on decimal point
\usepackage{bm}% bold math
\usepackage{color}
\definecolor{red}{rgb}{1,0.0,0} 
\definecolor{VPhighlight}{rgb}{0.1,0.6,0.9}
\definecolor{VPcomment}{rgb}{0,0,0.58}

\begin{document}

%\preprint{IFT-P.xx/2009}
%\preprint{ArXiv:yymm.nnnn}
\title{Dirac neutrinos in an $SU(2)$ left-right symmetric model 
%Dirac Neutrinos in two bi-doublets and one inert Higgs dublet left-right symmetric model
}

% \altaffiliation[Also at ]{Physics Department, XYZ University.}%

\author{ Henry Diaz}%
\email{hdiaz@uni.edu.pe}
\affiliation{Facultad de Ciencias, Universidad Nacional de Ingenier\'\i a (UNI), Av. Tupac Amaru s/n, R\'\i mac, Lima - 15333, Per\'u }

\author{V. Pleitez}%
\email{v.pleitez@unesp.br}
\affiliation{
Instituto  de F\'\i sica Te\'orica--Universidade Estadual Paulista \\
R. Dr. Bento Teobaldo Ferraz 271, Barra Funda\\ S\~ao Paulo - SP, 01140-070,
Brazil
}

\author{O. Pereyra Ravinez }%
\email{opereyra@uni.edu.pe}
\affiliation{Facultad de Ciencias, Universidad Nacional de Ingenier\'\i a (UNI), Av. Tupac Amaru s/n, R\'\i mac, Lima - 15333, Per\'u }

\date{10/08/2020}% It is always \today, today, versao de lunes 05/08
% sobre a versao do dia 12/12/2016
\begin{abstract}
We propose a left-right symmetric model, with the scalar sector consisting of two doublets and several bidoublets, 
in which neutrinos remain as Dirac fermions in all orders in perturbation theory. Although with only two bidoublets the neutrino masses still need a 
fine-tuning, this is not the case when a 
third bidoublet is added. One of the scalar doublets may be inert, since the left-right symmetry forbids it to 
couple with fermions.  
\end{abstract}


\maketitle

\section{Introduction}
\label{sec:intro}
According to the Standard Model (SM), neutrinos are massless particles. 
However, experiments~\cite{pdg2020}, indicate that neutrinos 
could have very small masses and that there is a mixing in the leptonic charged currents. However, the nature of neutrinos is still
unknown. Neutrinos may be: purely Majorana (equal to their charge 
conjugated fields), purely Dirac (different from their charge conjugated fields), quasi-Dirac (when two active Majorana neutrinos are mass 
degenerate), or pseudo-Dirac (when the mass degeneration occurs with one of them being active and the other one being a sterile neutrino); 
see~\cite{Machado:2010ui,Rossi-Torres:2013dya}, and references therein. In the last two cases, the mass degeneracy occurs at tree 
level but quantum corrections usually imply an additional small Majorana mass and, eventually neutrinos become Majorana particles.
In fact, it is difficult  to keep the lepton number $L$  automatically conserved in most extensions of the SM if
neutrinos are Majorana particles. 

However, considering them as Majorana particles, it is possible to explain the smallness of their masses, 
even at the tree level,  using the seesaw mechanism types I and II  in $SU(2)_L\otimes~U(1)_Y$ models, if complex ($Y=2$) 
scalar triplets and right-handed neutrinos $\nu_R$ are added~\cite{Schechter:1980gr,Cheng:1980qt}. These mechanisms can be
implemented, for instance, if new physics does exist at the TeV scale, in the context of models with
$SU(2)_L\otimes SU(2)_R\otimes U(1)_{B-L}$~\cite{Minkowski:1977sc}, in unified theories~\cite{gellmann}, and in
models with $SU(n)_F\otimes SU(2)\otimes U(1)_Y$ symmetries~\cite{Yanagida:1980xy}. Instead, the type III seesaw mechanism requires the
introduction of a self-conjugate ($Y=0$) triplet of fermions and can be implemented in  $SU(2)\otimes U(1)_Y$ or 
$SU(2)_L\otimes SU(2)_R\otimes U(1)_{B-L}$ models~\cite{Foot:1988aq}.

Among the best motivated extensions for the Electroweak Standard Model (ESM) are those with $SU(2)_L\otimes SU(2)_R\otimes U^\prime(1)$ gauge 
symmetry plus left-right parity symmetry~\cite{Pati:1974yy,Mohapatra:1974hk,Mohapatra:1974gc}, termed LR symmetric models. 
Although it is possible to introduce a generalized charge conjugation symmetry instead of a parity~\cite{Dekens:2014ina}, here we will consider 
only the case of parity. In particular, in these models the parity may be spontaneously broken~\cite{Senjanovic:1975rk}, and, moreover, the 
$U^\prime(1)$ factor can be identified with $B-L$, allowing one to implement quark and lepton correspondence since they are distinguished only
by the $B-L$ quantum number~\cite{Minkowski:1977sc, Davidson:1978pm,Marshak:1979fm}. One must bear in mind that this correspondence would be
stronger if neutrinos were Dirac particles. However, in the minimal LR model with one bidoublet and two doublets in the scalar sector the
smallness of the neutrino masses is not easily explained~\cite{Senjanovic:1978ev}. Then, in the context of this model, quarks and leptons are those 
of the SM plus three right-handed neutrinos which are incorporated naturally in a doublet, together with the right-handed charged leptons.

In LR symmetric models, Majorana neutrinos and the seesaw mechanism are obtained if, instead of scalar doublets 
$\chi_{L,R}$~\cite{Senjanovic:1978ev}, one introduces the scalar triplets $\Delta_{L,R}$ ~\cite{Minkowski:1977sc, Mohapatra:1979ia}. 
However, the true nature of neutrinos is still unkown whether they are purely Dirac or Majorana (with or without a seesaw mechanism).
Hence, one can ask, what would happen if the neutrinos are actually pure Dirac fermions? In addition, is it possible to have purely Dirac 
neutrinos if the only additional neutral fermions were right-handed neutrinos? After all, it would be interesting if the lepton-quark 
correspondence is maintained when all particles gain masses, but this implies that neutrinos have to be Dirac fermions.
In LR models with the scalar sector consisting of only one bidoublet and two doublets~\cite{Senjanovic:1978ev} it is possible to accommodate Dirac neutrino masses.

Models with Dirac neutrinos have also been proposed, where the smallness of their masses is explained. For instance, the calculable Dirac
neutrino masses were obtained in the context of LR symmetric models by introducing extra heavy singlet 
leptons and/or charged and neutral scalars~\cite{Chang:1986bp,Mohapatra:1987hh,Borah:2017leo}, or even doubly charged scalars \cite{Ma:2017kgb}. 
Moreover, recently~\cite{Ma:2015raa}, anomaly-free models were proposed that allow Dirac or inverse seesaw neutrino masses, which include sterile neutrinos with 
exotic lepton number assignment~\cite{Montero:2007cd}. To obtain calculable Dirac neutrino masses in the scotogenic models we have to add, for instance,
besides the right-handed neutrino, (i) two neutral leptons $N$ and $N^c$ per family or (ii) two new fermion singlets and one fermion 
doublet~\cite{Ma:2019yfo}. It is also possible to implement the inverse seewsaw mechanism without the introduction of triplets, but in such a case
one must add more neutral singlet leptons~\cite{Brdar:2018sbk}. Recently,   an alternative formulation of the LR symmetric
models has been proposed in which $B-L$ is a global unbroken symmetry and where neutrinos are Dirac particles~\cite{Bolton:2019bou}; 
the price to be paid is the introduction of extra quarks and charged leptons. 
It means that, with only the known leptons plus three right-handed neutrinos, and using only renormalizable interactions, purely massive Dirac 
neutrinos do not arise easily in any model. Hence, it is interesting to search mechanisms that allow one to accommodate light Dirac neutrinos in the context of a renormalizable electroweak model with a representation content in a complete analogy with the SM regarding the charged fermions, the only extra neutral fermions being three right-handed neutrinos. 

On the other hand, although the resonance discovery at the LHC~\cite{Chatrchyan:2012ufa,Aad:2012tfa} is consistent with the neutral scalar of the SM, 
it does not discard the existence of more neutral scalars (and their charged partners if these scalars are not singlets of the gauge symmetry). 
Since the scalar content in any model is not fixed by the gauge symmetry, and also we do not know yet the complete spectra in the scalar sector, 
we can add, in any model, more  scalar multiplets.
Hence, the issue of the number of scalars is added to the generation problem: How many scalars? The interesting possibility is that this 
number is equal to the number of fermion generations, i.e., three~\cite{Machado:2012ed}. Although in the context of the SM, the introduction of 
three doublets is well motivated, e.g., for implementing $C\!P$ violation~\cite{Weinberg:1976hu} and/or dark matter candidates~\cite{Fortes:2014dca}, 
in models with larger gauge symmetries a given number of scalar multiplets is not necessarily well motivated.  This depends on the 
phenomenological results. This is the case in the LR symmetric electroweak models in which there are several ways to introduce scalar 
multiplets. Here, we will consider an extension of the minimal LR model by adding more bidoublets and no triplets.  In particular we 
show that in the case of three bidoublets it is possible to avoid a 
fine-tuning in the neutrino Yukawa couplings. However, the details
of the scalar potential are given only for the case of two bidoublets and two doublets.  
 
The outline of this work is as follows. In the next section we consider the model and the symmetries that make it invariant under a generalized
parity and other discrete symmetries in such a way that one bidoublet is in one case coupled only with leptons and the other only with quarks. In 
Sec.~\ref{sec:potential} we consider the most general scalar potential invariant under the symmetries of the model. We show that for the 
two bidoublets case an approximate  $\mathbb{Z}_5$ symmetry allows us to consider a more simplified potential. The gauge vector boson sector
is analyzed in Sec.~\ref{sec:gaugebosons}, while Yukawa interactions and fermion masses are considered in Sec.~\ref{sec:yukawa}. Next, 
in Sec.~\ref{sec:leptonsgauge} the fermion-vector boson interactions are given, while in Sec.~\ref{sec:lmasses} we analyze the case when
we add a third bidoublet. Some phenomenological consequences are described in Sec.~\ref{sec:feno}, and, in Sec.~\ref{sec:parityfirst} the case where
parity is broken first is considered. Finally our conclusions appear in the last section.

\section{The model}
\label{sec:model}

The model to be considered has the following electroweak symmetry:
\begin{equation}
SU(2)_L\otimes SU(2)_R\otimes U(1)_{B-L}\otimes\mathcal{P},
\label{simmetry}
\end{equation}
We omit the $SU(3)_C$ factor because is similar to the SM. The electric charge operator is defined as usual: $Q= T_{3L}+T_{3R}+(B-L)/2$.

The left- and right-handed fermions transform nontrivially under different $SU(2)$ transformations. In the lepton sector $L^{\prime T}_l =(\nu^\prime_l\, l^\prime)_L\sim(\textbf{2}_L,\textbf{1}_R,-1)$ and $R^{\prime T}_l =(\nu^\prime_l\, l^\prime)_R\sim(\textbf{1}_L,\textbf{2}_R,-1)$,
with $l=e,\mu,\tau$ and the primed states denote symmetry eigenstates. Si\-mi\-lar\-ly, in the quark sector,
$Q^{\prime T}_L\sim(\textbf{2}_L,\textbf{1}_R,-1/3)$ and $Q^{\prime T}_R\sim(\textbf{1}_L,\textbf{2}_R,-1/3)$.
The scalar sector consists of at least two or three bidoublets $\Phi_i$ transforming as $(\textbf{2}_L,\textbf{2}^*_R,0)$ and two doublets $\chi^T_L=(\chi^+_L\;\chi^0_L)\sim(\textbf{2}_L,\textbf{1}_R,+1)$ and  $\chi^T_R=(\chi^+_R\;\chi^0_R)\sim(\textbf{1}_L,\textbf{2}_R,+1)$ to break the parity and the gauge symmetry down to $U(1)_Q$~\cite{Senjanovic:1975rk,Senjanovic:1978ev}. 

We also impose a generalized parity under which  
\begin{equation}
g_L\leftrightarrow g_R ,\;W_{L\mu}\leftrightarrow W^\mu_{R},\; f_L\leftrightarrow f_R,\;\chi_L\leftrightarrow \chi_R, \Phi_i\leftrightarrow \Phi^\dagger_i,\tilde{\Phi}_i\leftrightarrow \tilde{\Phi}^\dagger_i,
\label{lr}
\end{equation}
where $\tilde{\Phi}_i=\tau_2\Phi^*_i\tau_2$; $W_{\mu L,R}$ are the vectorial gauge bosons of the $SU(2)_{L,R}$ gauge symmetry, 
respectively, $f$ denotes a quark or a lepton doublet, and $\Phi_i$ and $\chi_{L,R}$ are the scalar multiplets  introduced above. 
The coupling constants $g_{L,R},g^\prime$ correspond to the groups $SU(2)_{L,R}$ and $U(1)_{B-L}$, respectively. However, the 
invariance under $\mathcal{P}$ implies equality of gauge couplings $g_L=g_R\equiv g$ at the energy at which these symmetries are 
realized. Under this condition the model has only two gauge couplings, $g$ and $g^\prime$. Although, as a result of running
couplings, we will have $g_L \not= g_R$~\cite{Chang:1984uy}, we  consider in this paper the case when these two couplings are equal at
any energy scale but this has to be seen just as an approximation.

For the case of two bidoublets, we will impose also the discrete symmetries $\mathbb{Z}_2\times \mathbb{Z}^\prime_2$, in such a 
way that under the first factor $L_R,\Phi_1\to -L_R,-\Phi_1$, and under the second one  $Q_R,\Phi_2\to -Q_R,-\Phi_2$, while all
the other fields transform trivially under both factors. This symmetry implies that the bidoublet $\Phi_1$ couples only to leptons
and the other, $\Phi_2$, only to quarks. Notice that, as usual in this sort of models, as a consequence of the transformation under
the $SU(2)_{L,R}$ factors, none of the scalar doublets couple to the fermions.

\section{The scalar potential}
\label{sec:potential}

First, we are going to consider the most general scalar potential, invariant under the gauge symmetries and parity, and next we see the effect 
of imposing discrete symmetries. Since some of our results are valid for an arbitrary number of bidoublets we consider the scalar potential 
involving $n$ bidoublets and two doublets.
In general, a bidoublet $\Phi$ transforms under the $SU(2)_L\otimes SU(2)_R$ symmetry as 
$\Phi\to U_L\Phi U^\dagger_R$, $\Phi^\dagger\to U_R\Phi^\dagger U^\dagger_L$ and
$\tilde{\Phi}\to U_L\tilde{\Phi}U^\dagger_R$, $\tilde{\Phi}^\dagger\to U_R\tilde{\Phi}^\dagger U^\dagger_L$. Under 
these conditions, the scalar potential is given by
\begin{equation}
V=V^{(2)}+V^{(4a)}+V^{(4b)}+V^{(4c)}+V^{(d)}, 
\label{potential0}
\end{equation}
where
\begin{eqnarray}
V^{(2)}&=&\frac{1}{2}\sum_{i,j=1}^n\left[\mu^2_{ij}\textrm{Tr}(\Phi^\dagger_i\Phi_j)+\tilde{\mu}^2_{ij}\textrm{Tr}(\tilde{\Phi}^\dagger_i\Phi_j)+H.c.\right]+
\mu^2_{LR}(\chi^\dagger_L\chi_L+\chi^\dagger_R\chi_R),
\nonumber \\
V^{(4a)}&=&  \frac{1}{2}\sum_{i,j=1}^n\left[
\lambda_{ij} \textrm{Tr}(\Phi^\dagger_i\Phi_j)^2 +\tilde{\lambda}_{ij}\textrm{Tr} (\tilde{\Phi}^\dagger_i\Phi_j)^2
+H.c.\right],
\nonumber \\
V^{(4b)}&=&
\frac{1}{2}\left[\sum_{i,j=1}^n\lambda^\prime_{ij}(\textrm{Tr}\Phi^\dagger_i\Phi_j)^2+\tilde{\lambda}^\prime_{ij}(\textrm{Tr}\tilde{\Phi}^\dagger_i\Phi_j)^2+H.c.\right],
\nonumber  \\
V^{(4c)}&=&\sum_{ij}[\rho_{ij}\textrm{Tr}(\Phi^\dagger_i\Phi_i\Phi^\dagger_j\Phi_j)
+\tilde{\rho}_{ij}\textrm{Tr}(\tilde{\Phi}^\dagger_i\Phi_i\tilde{\Phi}^\dagger_j\Phi_j)],
\nonumber \\
V^{(4d)}&=&
\frac{1}{2}[\sum_{i,j=1}^n(\Lambda_{ij}\textrm{Tr}\Phi^\dagger_i\Phi_j+
\tilde{\Lambda}_{ij}\textrm{Tr}\tilde{\Phi}^\dagger_i\Phi_j)
(\chi^\dagger_L\chi_L+
\chi^\dagger_R\chi_R)\nonumber \\ &+&  \bar{\Lambda}_{ij}(\chi^\dagger_L\Phi_i\Phi^\dagger_j\chi_L+\chi^\dagger_R\Phi^\dagger_i\Phi_j\chi_R)+
\Omega_{ij}(\chi^\dagger_L\tilde{\Phi}_i\Phi^\dagger_j\chi_L+\chi^\dagger_R\tilde{\Phi}^\dagger_i\Phi_j\chi_R)
\nonumber \\&+&
\bar{\Lambda}^\prime_{ij}(\chi^\dagger_L\tilde{\Phi}_i\tilde{\Phi}^\dagger_j\chi_L+
\chi^\dagger_R\tilde{\Phi}^\dagger_i\tilde{\Phi}_j\chi_R) +\Omega^\prime_{ij}(\chi^\dagger_L\Phi_i\tilde{\Phi}^\dagger_j\chi_L+
\chi^\dagger_R\Phi_i\tilde{\Phi}^\dagger_j\chi_R) +H.c ],
\nonumber \\ 
V^{(4e)}&=&\lambda_{LR}[(\chi^\dagger_L\chi_L)^2+(\chi^\dagger_R\chi_R)^2].
\label{potencial1}
\end{eqnarray}
We have omitted the redundant terms, for instance $\textrm{Tr}(\Phi^\dagger_i\Phi_j)=\textrm{Tr}(\tilde{\Phi}^\dagger_i\tilde{\Phi}_j)$, and so on. 

Let us consider explicitly the case of two bidoublets, $n=1,2$ in (\ref{potencial1}) with 
\begin{equation}
\Phi_1 = \left(\begin{array}{cc}
\phi^{0}_{1} & \eta^{+}_{1}\\
\phi^{-}_{1} & \eta^{0}_{1}
\end{array}\right),\;\Phi_2 = \left(\begin{array}{cc}
\phi^{0}_{2} & \eta^{+}_{2}\\
\phi^{-}_{2} & \eta^{0}_{2}
\end{array}\right).
\label{higgs1}
\end{equation}

The vacuum expectation values (VEVs) are $\sqrt{2}\langle\Phi_1\rangle\!\!=\!\!\textrm{Diag}(k_1\,k^\prime_1)$,  $\sqrt{2}\langle\Phi_2\rangle~=~\textrm{Diag}(k_2\,k^\prime_2)$, $\sqrt{2}\langle\chi_L\rangle = \textrm{Diag}(0\,v_L)$, and $\sqrt{2}\langle\chi_R\rangle = \textrm{Diag}(0\,v_R)$.
In general, we will write the neutral components of the scalars as $ x^0_i=\frac{1}{\sqrt2}(v_i+R_i+iI_i)e^{i\theta_i} $ ,
where $v_i=k_i,k^\prime_i,\;i=1,2$; and $v_L$ and $v_R$ may be complex numbers and $R_i$ and $I_i$, Hermitian fields. However, here we will consider all VEVs real for all $i$ running over the bi-doublets and doublets.

In this case,  the invariance under the parity transformations defined in (\ref{lr}) implies~$\mu_{12}~=~\mu_{21}\equiv~\mu^2$, 
$\tilde{\mu}_{12}~=~\tilde{\mu}_{21}\equiv~\nu^2$; $\lambda_{12}=\lambda_{21},\;
\;\tilde{\lambda}_{12}=\tilde{\lambda}_{21},\;\lambda^\prime_{12}=\lambda^\prime_{21},\;
\Lambda_{12}=\Lambda_{21},\; \bar{\Lambda}_{12}=\bar{\Lambda}_{21},\;\bar{\Lambda}^\prime_{12}=\bar{\Lambda}^\prime_{21},  
$ and that $\tilde{\lambda}_{ij}$ and $\Omega_{ij}$ are real. Notice that the  $\mathbb{Z}_2\otimes \mathbb{Z}^\prime_2$ implies $\mu^2=\nu^2=0$ and $\Lambda_{12}=\Lambda_{21}= \bar{\Lambda}_{12}=\bar{\Lambda}_{21}=\bar{\Lambda}^\prime_{12}=\bar{\Lambda}^\prime_{21}=0$. 
However, we will allow for the moment a soft breaking of these symmetries and use $\mu^2\not=0$.

The constraint equations $t_i=\partial V/\partial X_i,\;X_i=k_1,k^\prime_1,k_2,k^\prime_2,v_L,v_R$ (considered real), are
\begin{eqnarray}
t_1&=&k_1\left[\mu^2_{11}+\left(\lambda_{11}+\lambda^{\prime}_{11}\right)k^2_1+ k^{\prime 2}_1\,\left(\lambda^{\prime}_{11}+
\tilde{\lambda}_{11}+2\tilde{\lambda}^{\prime}_{11}\right)+\frac{1}{2}\left(v^2_RH+
\tilde{\lambda}_{21}k^{\prime 2}_2\right.\right.\nonumber \\
&+&\left.\left.\left (\tilde{\lambda}_{12}+\tilde{\lambda}^{\prime}_{21}+\tilde{\lambda}^{\prime}_{12}\right)k^{\prime 2}_2+
\frac{1}{2}v^2_LH+k^2_2\left(\lambda^{\prime}_{12}+\lambda^{\prime}_{21}+\lambda_{21}+\lambda_{12}+\rho_{12}\right)\right)\right] \nonumber  \\
&+&\frac{1}{4}(v^2_R + v^2_L)(k^{\prime}_1 D  +k^{\prime}_2  F+k_2 G )
+\frac{k_2k^{\prime}_1 k^{\prime}_2}{2}\left(\tilde{\lambda}^{\prime}_{21}+\tilde{\lambda}^{\prime}_{12}+\lambda^{\prime}_{12}+\lambda^{\prime}_{21}+\tilde{\rho}_{12}\right)\nonumber \\ &+&
k_2\mu^2+\tilde{\mu}^2_{11}k^{\prime}_1,
\label{vin11}
\end{eqnarray}

\begin{eqnarray}
t^{\prime}_1&=&k^\prime_1\left[\mu^2_{11}+\left(\lambda_{11}+\lambda^{\prime}_{11}\right)k^{\prime 2}_1+k^{2}_1\,\left(\lambda^{\prime}_{11}+\tilde{\lambda}_{11}+
2\,\tilde{\lambda}^{\prime}_{11}\right)+\frac{1}{2}(v^2_R H+
\tilde{\lambda}^{\prime}_{12}k^{2}_2\right. \nonumber \\
&+&\left.\left.(\tilde{\lambda}^{\prime}_{21}+\tilde{\lambda}_{12}+\tilde{\lambda}_{21})k^{2}_2+\frac{1}{2}
v^2_LH+k^{\prime 2}_2\left(\lambda^{\prime}_{12}+\lambda^{\prime}_{21}+\lambda_{21}+\lambda_{12}+\rho_{12}\right)\right) \right]\nonumber \\
&+&\frac{1}{4} \left (v^2_L+v^2_R\right)(k_1D +
k_2 F +k^{\prime}_2 G )+
\frac{k_2 k_1 k^{\prime}_2}{2}\left(\tilde{\lambda}^{\prime}_{12}+\tilde{\lambda}^{\prime}_{21}+\lambda^{\prime}_{12}+\lambda^{\prime}_{21}+\tilde{\rho}_{12}\right)
\nonumber \\&+&k^{\prime}_2\mu^2+\tilde{\mu}^2_{11}k_1,
\label{vin12}
\end{eqnarray}
%where 
%\begin{eqnarray}
%&&A= \Lambda_{11}+\overline{\Lambda}_{11},\,\, B=\Lambda_{12}+\Lambda_{21}+\overline{\Lambda}_{12}+\overline{\Lambda}_{21},\,\, C=\tilde{\Lambda}_{12}+\tilde{\Lambda}_{21}+\Omega^{\prime}_{12},
%+\Omega_{21}\nonumber  \\&& D= \Omega^{\prime}_{11}+2\,\tilde{\Lambda}_{11}+\Omega_{11},\,\, E=\Omega^{\prime}_{22}+2\,\tilde{\Lambda}_{22}+\Omega_{22},\,\, F=\Omega^{\prime}_{21}+\tilde{\Lambda}_{21}+\tilde{\Lambda}_{12}
%+\Omega_{12},\nonumber \\&&  G=\overline{\Lambda}^{\prime}_{21}+\overline{\Lambda}^{\prime}_{12}+\Lambda_{21}+\Lambda_{12},\,\, H=\Lambda_{11}+\overline{\Lambda}^{\prime}_{11},
%\nonumber \\ && I=\Lambda_{22}+\overline{\Lambda}^{\prime}_{22},\,\, J=\Lambda_{22}+\overline{\Lambda}_{22},
%\label{defx}
%\end{eqnarray}
and similarly we obtain $t_2$ and $t^\prime_2$ for $k_2$ and $k^\prime_2$, respectively, but we will not write them explicitly. Finally, we have
\begin{equation}
t_L=\frac{v_L}{2}(2\mu^2_{LR}+2\lambda_{LR}v^2_L+\Delta),\quad 
t_R=\frac{v_R}{2}(2\mu^2_{LR}+2\lambda_{LR}v^2_{R}+\Delta),
\label{tlr}
\end{equation}
where
\begin{eqnarray}
\Delta&=&k^{\prime 2}_1 A +
k^{\prime}_1k^{\prime}_2B +
k^{\prime}_1k_2 C +k_1k^{\prime}_1  D +
k_2k^{\prime}_2 E  +k_1k^{\prime}_2 F 
+k_1k_2 G+k^{2}_1 H +k^2_2 I  \nonumber \\&+&
k^{\prime 2}_2 J,
\label{LR}
\end{eqnarray}
and
\begin{eqnarray}
A&=& \Lambda_{11}+\overline{\Lambda}_{11},\,\, B=\Lambda_{12}+\Lambda_{21}+\overline{\Lambda}_{12}+\overline{\Lambda}_{21},\,\, C=\tilde{\Lambda}_{12}+\tilde{\Lambda}_{21}+\Omega^{\prime}_{12},
+\Omega_{21}\nonumber  \\ D&=& \Omega^{\prime}_{11}+2\,\tilde{\Lambda}_{11}+\Omega_{11},\,\, E=\Omega^{\prime}_{22}+2\,\tilde{\Lambda}_{22}+\Omega_{22},\,\, F=\Omega^{\prime}_{21}+\tilde{\Lambda}_{21}+\tilde{\Lambda}_{12}
+\Omega_{12},\nonumber \\  G&=&\overline{\Lambda}^{\prime}_{21}+\overline{\Lambda}^{\prime}_{12}+\Lambda_{21}+\Lambda_{12},\,\, H=\Lambda_{11}+\overline{\Lambda}^{\prime}_{11},
\nonumber \\  I&=&\Lambda_{22}+\overline{\Lambda}^{\prime}_{22},\,\, J=\Lambda_{22}+\overline{\Lambda}_{22},
\label{defx}
\end{eqnarray}
Notice that only $v_L$ and $v_R$ can be zero; however, this solution is not accepted for $v_R$.
We assume also that $v_R\gg k_2\gg k_1,k^\prime_1,k_2^\prime\gg v_L$, and if  
\begin{eqnarray}
&& D,F,G\ll1,\;\;
%\nonumber \\&&
\tilde{\lambda}^{\prime}_{12}+\tilde{\lambda}^{\prime}_{21}+\lambda_{21}+\lambda_{12}+\tilde{\rho}_{12}\ll1,
\label{app}
\end{eqnarray}
then we obtain from Eqs.~(\ref{vin11}) and (\ref{vin12}), respectively, 
\begin{equation}
k_1\approx \frac{\mu^2}{\mu^2_{11}+v^2_RH}\,k_2\ll k_2,\qquad k^\prime_1\approx 
\frac{\mu^2}{\mu^2_{11}+v^2_RH}\,k^\prime_2\ll k^\prime_2,
\label{t1}
\end{equation}
with $v ^2_RH> \vert \mu^2_{11}\vert$. This shows that there is a range of the parameter space in which we can have $k^\prime_1\ll k_1\ll k^\prime_2<k_2$. Moreover, if we assume
\begin{equation}
D=F=G=0,\quad \lambda_{ij}=\lambda^\prime_{ij}=\tilde{\lambda}_{ij}=\tilde{\lambda}^\prime_{ij}=\tilde{\rho}_{ij}=0,\;i\not= j;\;\mu^2=\tilde{\mu}^2_{11}=0,
\label{vin3}
\end{equation}
the constraint equations become 
\begin{eqnarray}
&& t_1=k_1\left(\mu^2_{11} +(\lambda_{11}+\lambda^\prime_{11})k^2_1+\lambda^\prime_{11}k^{\prime2}_1+\frac{1} {2}(v^2_L+v^2_R)H+
\frac{1} {2}\rho_{12}k^2_2\right),
\nonumber \\ &&
%\begin{eqnarray}
t^\prime_1=k^\prime_1\left(\mu^2_{11}+\lambda^\prime_{11}k^2_1+(\lambda_{11}+\lambda^{\prime2}_{11})
k^{\prime2}_1+\frac{1}{2}(v^2 _L+v^2_R)A+\frac{1} {2}\rho_{12}k^{\prime2}_2
\right),\nonumber \\&&
t_2=k_2\left(\mu^2_{22}+(\lambda_{22}+\lambda^\prime_{22})k^2_2+\lambda^\prime_{22}k^{\prime2}_2+
\frac{1}{2}(v^2_L+v^2_R) I+\frac{1} {2}\rho_{12}k^2_1\right) ,
\nonumber \\&&
t^\prime_2=k^\prime_2\left(\mu^2_{22}+(\lambda_{22}+\lambda^\prime_{22})k^{\prime2}_2+\lambda^\prime_{22}k^2_2+ \frac{1}{2}(v^2_L+v^2_R)J+\frac{1} {2}\rho_{12}k^{\prime2}_1\right),
\nonumber \\&&
t_L=\frac{v_L}{2}\left[2\mu^2_{LR}+2\lambda_{LR}v^2_L+k^{\prime2}_1A
+k^2_1H
+k^2_2I +k^{\prime2}_2J\right],
\nonumber \\&&
t_R=\frac{v_R}{2}\left[2\mu^2_{LR}+2\lambda_{LR}v^2_R+ k^{\prime2}_1A
+k^2_1H+k^2_2I +k^{\prime2}_2J\right].
\label{vinculos}
\end{eqnarray}

In fact, we further restrict the Higgs potential so that it is invariant under the  $\mathbb{Z}_5$ symmetry
(defined as  $\omega_i=e^{2\pi\,i \,n/5},\; n = 0,\cdots, 4$) under
which $\Phi_1\to \omega_1\Phi_1,\;\Phi_2\to \omega_2\Phi_2$ while also other fields are invariant; in this way, the scalar potential 
in Eq.~(\ref{potencial1}) becomes
\begin{eqnarray}
V^{(2)}&=&\frac{1}{2}\sum_{i=1,2}\left[\mu^2_{ii}\textrm{Tr}(\Phi^\dagger_i\Phi_i)+
H.c.\right]+
\mu^2_{LR}(\chi^\dagger_L\chi_L+\chi^\dagger_R\chi_R),
\nonumber \\
V^{(4a)}&=&  \frac{1}{2}\sum_{i=1,2}\left[
\lambda_{ii} \textrm{Tr}(\Phi^\dagger_i\Phi_i)^2
+H.c.\right],
\nonumber \\
V^{(4b)}&=&
\frac{1}{2}\sum_{i=1,2}\lambda^\prime_{ii}(\textrm{Tr}\Phi^\dagger_i\Phi_i)^2,
\nonumber  \\
V^{(4c)}&=&\rho_{12}\textrm{Tr}(\Phi^\dagger_1\Phi_1\Phi^\dagger_2\Phi_2),
\nonumber \\
V^{(4d)}&=&
\frac{1}{2}\left[\sum_{i=1,2}\left\{\Lambda_{ii}\textrm{Tr}\Phi^\dagger_i\Phi_i(\chi^\dagger_L\chi_L+\chi^\dagger_R\chi_R)
+ \bar{\Lambda}_{ii}(\chi^\dagger_L\Phi_i\Phi^\dagger_i\chi_L+\chi^\dagger_R\Phi_i\Phi^\dagger_i\chi_R)+\right. \right.\nonumber \\
&+&\left. \left.\bar{\Lambda}^\prime_{ii}(\chi^\dagger_L\tilde{\Phi}_i\tilde{\Phi}^\dagger_i\chi_L+
\chi^\dagger_R\tilde{\Phi}_i\tilde{\Phi}^\dagger_i\chi_R)\right\}\right] ,
\nonumber \\ 
V^{(4e)}&=&\lambda_{LR}[(\chi^\dagger_L\chi_L)^2+(\chi^\dagger_R\chi_R)^2],
\label{potencial2}
\end{eqnarray}
and the constraints in Eq.~(\ref{vin3}) arise from this potential. It means that these conditions are protected by the 
$\mathbb{Z}_5$ symmetry and may be naturally small. We may consider the potential in Eq.~(\ref{potencial2}), and the respective 
mass spectra, as a good approximation. Notice that  all VEVs may be zero; in particular, the solutions $k^\prime_{1,2}=0$ and $v_L=0$ 
are allowed. The SM-like Higgs scalar is in the bidoublet $\Phi_2$.

It is important to note that since the doublet $\chi_L$ was introduced just to implement the invariance of the Lagrangian under 
parity and it does not couple to fermions, if the respective VEV is zero it becomes an inert doublet since the left-right symmetry 
protects its inert character; hence it is a candidate for dark matter.  However, notice that $v_L\not=0$ is also a solution, hence 
the possibility to have a model without any bidoublet, with fermion masses arisen from nonrenormalizable 
interactions~\cite{Brahmachari:2003wv}; in this case, it is possible to make $A=H=I=J=0$ in Eq.~(\ref{vinculos}). However, in this 
case, the model needs an ultraviolet completion. We stress that although the constraint equations in Eq.~(\ref{vinculos}) were
obtained using the potential in Eq.~(\ref{potencial2}) 
by considering the most general potential (without the $\mathbb{Z}_2\otimes \mathbb{Z}^\prime_2$ symmetries), we still obtain
\begin{equation}
t_L=v_L(\mu^2_{LR}+\lambda_{LR}v^2_L+\textrm{bidoublet contributions}),
\label{vinculos2}
\end{equation}
and the solution $v_L=0$ is still allowed even without a soft breaking of parity symmetry~\cite{Siringo:2004hm}.

If we denote $x^0=\phi^0_{1,2},\eta^0_{1,2}$ (we omit the respective VEV) and we take the symmetry eigenstate as $x^0_i=R_i+i I_i$, they 
are related with the mass eigenstates, $H^0$ and $A^0$, throughout the orthogonal $4\times4$ matrices, say $R_i=O_{ij}H_j$ and 
$I_i=\mathcal{O}_{ij}A_j$. Recall that we are not assuming $CP$ violation; hence, $O$ and $\mathcal{O}$ are orthogonal matrices. 

\section{Gauge bosons mass eigenstates}
\label{sec:gaugebosons}

The covariant derivative for the bidoublets $\Phi_i,\;i=1,2$, and for the doublets $\chi_L$ and $\chi_R$ are given by 
\begin{eqnarray}
\mathcal{D}_{\mu }\Phi_i&=&\partial _{\mu }\Phi_i+ig\left[\frac{\vec{\tau}}{2}\cdot\vec{W}_L\Phi_i-\Phi_i  \frac{\vec{\tau}}{2}\cdot\vec{W}_R\right],\quad (a)\nonumber \\
\mathcal{D}_{\mu }\chi_L&=&\left(\partial _{\mu }+ig \frac{\vec{\tau}}{2}\cdot \vec{W}_L+g^\prime B_{\mu }\right)\chi_L, \quad \quad\quad(b)\nonumber \\ 
\mathcal{D}_{\mu }\chi_R&=&\left(\partial _{\mu }+ig \frac{\vec{\tau}}{2}\cdot \vec{W}_R+g^\prime B_{\mu }\right)\chi_R,\quad \quad \quad (c)
\label{dcovariante}
\end{eqnarray}
where we have already established $g_L=g_R=g$ (see Sec.~\ref{sec:model}). With the VEVs given in Sec.~\ref{sec:model} we obtain 
for the charged vector bosons
\begin{equation}
M^2_{CB}=\frac{g^2v^2_R}{4}\left(\begin{array}{cc}  %A=K, B=K^\prime
x+y& -2z \\
&1+x 
\end{array} 
\right), 
\label{massws1}
\end{equation}
where
\begin{eqnarray*}
   x&=&K^2/v^2_R,\hspace{0.7cm} z=\bar{K}^2/v^2_R,\hspace{0.7cm} y=v^2_L/v^2_R,\\
  K^2&=&k^2_1+k^{\prime2}_1+k^2_2+k^{\prime2}_2,\hspace{0.4cm}\bar{K}^2=k_1k^\prime_1+k_ 2k^\prime_2,
\end{eqnarray*}
and the respective eigenvalues are given by
\begin{equation}
M^2_{W_1}=\frac{g^2v^2_R}{4}\left(x+\frac{1+y}{2}-\sqrt{\Delta}\right),\quad
M^2_{W_2}=\frac{g^2}{4}\left(x+\frac{1+y}{2}+\sqrt{\Delta}\right),
\label{massws2}
\end{equation}
where $\Delta=4z+\frac{1}{4}(y-1)^2$. These expressions can be generalized for an arbitrary number of bidoublets $K^2=\sum_i^n k^2_i$, and $\bar{K}^2=\sum_i^n k_ik^\prime_i$ and the results of this section are valid for $n$ bidoublets.

Mass and symmetry vector eigenstates are related by an orthogonal matrix
\begin{equation}
\left(\begin{array}{c}
W^+_{1\mu}\\
W^+_{2\mu}
\end{array} 
\right) =\left(\begin{array}{cc}
c_\xi & s_\xi\\
-s_\xi& c_\xi
\end{array} 
\right)\left(\begin{array}{c}
W^+_{L\mu}\\
W^+_{R\mu}
\end{array} 
\right),
\label{ws1}
\end{equation}
with
\begin{equation}
c_\xi=\frac{Y}{\sqrt{16z+Y^2}},\;s_\xi=\frac{\sqrt{16z}}{\sqrt{16z+Y^2}},
\label{ws2}
\end{equation}
where $Y=1-y+2\sqrt{\Delta}$ and $c_\xi=\cos\xi$, etc. Notice that since we will always consider that $k_{1,2},k^\prime_{1,2}\ll v_R$ i.e., $z\ll Y$, the mixing angle between $W^+_{L\mu}-W^+_{R\mu}$ can be arbitrarily small.

Note that $M^2_{W2}\gg M^2_{W1}$, and then we identify the $W^\pm$ of the Standard Model with $W^\pm_1$. In the limit $v_R\to \infty$ ($x,z,y\ll1$), we obtain 
\begin{equation}
M^2_{W_1}\approx \frac{g^2}{4}\left(K^2+\frac{v^2_L}{2}\right),\quad
M^2_{W_2}\approx \frac{g^2}{4}v^2_R.
\label{massws22}
\end{equation}

In the neutral vector bosons, we have the mass matrix:
\begin{equation}
M^2_{NB}=\frac{g^2v^2_R}{4}\left(\begin{array}{ccc}
x+y&-x  & -ry\\
& 1+x & -r\\
& & r^2(1+y)
\end{array} 
\right) ,
\label{masszs1}
\end{equation}
where we have defined $x$ and $y$ as before and $r\equiv g^\prime/g$. 

The determinant of the matrix in (\ref{masszs1}) is zero and its eigenvalues are without any approximation
\begin{eqnarray}
&& M_A=0,\nonumber \\&&
M^2_{Z_1}=\frac{g^2v^2_R}{4}\left[x+\frac{1}{2}(1+r^2)(1+y )-\frac{1}{2}\sqrt{\Omega} \right],\nonumber \\&&
M^2_{Z_2}=\frac{g^2v^2_R}{4}\left[x+\frac{1}{2}(1+r^2)(1+y)+\frac{1}{2}\sqrt{\Omega}\right],
\label{masszs2}
\end{eqnarray}
where we have defined
\begin{equation}
\Omega=(1+r^2)^2(1+y)^2-4(1+2r^2)y-4xr^2\,(1+y^2)+4x^2.
\label{zs1}
\end{equation}
%recall that $r\equiv r_\theta=g^\prime/g$ is given in Eq.~(\ref{tan1}) in terms of the angle $\theta$. 

The symmetry eigenstates ($W_{3L},W_{3R}$ and $B$) are linear combinations of the mass eigentates ($A,Z_1$ and $Z_2$) as follows:
\begin{equation}
\left(\begin{array}{c}
W_{3L}\\ W_{3R}\\ B
\end{array}\right)=
\left(\begin{array}{ccc}
n& n_{12} & n_{13} \\
n & n_{22} & n_{23} \\
n^\prime & n_{32} & n_{33}
\end{array}\right)
\left(\begin{array}{c}
A\\ Z_1 \\ Z_2
\end{array}\right)
\label{nb1}
\end{equation}
Although we have all the elements of $n_{ij}$ exactly calculated, here we write for the sake of space only $n$ and $n^ \prime$ 
exactly, while the other entries are in the approximation $v_R\gg K,v_L$  ($x,y,z\ll1$) up to $\mathcal{O}(1/v^2_R)$ terms:
\begin{eqnarray}\label{almost}
n&=&\sin\theta,\hspace{4.2cm}n^\prime=\sqrt{\cos2\theta},\hspace{1cm} n_{12}\approx\cos\theta,\hspace{1cm}
n_{13}\approx\phi,\nonumber \\ 
n_{22}&\approx& - t_\theta s_\theta 
\left(1-\phi\,\frac{\sqrt{c_{2\theta} }}{s^2_\theta c_\theta}\right) ,\hspace{1.1cm}
n_{23}\approx\frac{\sqrt{c_2\theta}}{c_\theta}\left(1+\phi \,\frac{c_\theta t^2_\theta}{\sqrt{c_{2\theta}}}\right),\nonumber
\\
n_{32}&\approx& 
-t_\theta  \sqrt{c_{2\theta}}\left(1+\phi\,\frac{1}{c_\theta\sqrt{c_{2\theta}}}\right),\hspace{0.34cm}
n_{33}\approx -t_\theta\left(1-\phi\,\frac{\sqrt{c_{2\theta}}} {c_\theta}\right),
\end{eqnarray}
where we have defined $ \phi=(x-r^2y)/(1+r^2)^{3/2} $. The angle $\theta$ is defined below:
 \begin{equation}
g=\frac{e}{\sin\theta},\quad g^\prime =\frac{e}{\sqrt{\cos2\theta}}.
\label{charge}
\end{equation}

In the limit $v_R\to\infty$ i.e., $x,y\to 0$ ($\phi\to0$, also), the matrix in Eq.~(\ref{nb1}) becomes the usual form in the literature:
\begin{equation}
\left(\begin{array}{c}
W_{3L}\\W_{3R}\\ B
\end{array}\right)=
\left(\begin{array}{ccc}
s_\theta & c_\theta & 0 \\
s_\theta &-s_\theta t_\theta & -\frac{\sqrt{c_{2\theta}}}{c_\theta} \\
\sqrt{c_{2\theta}} &-t_\theta \sqrt{c_{2\theta}}& -t_\theta
\end{array}\right)
\left(\begin{array}{c}
A \\ Z_1 \\ Z_2
\end{array}\right).
\label{nb222}
\end{equation}

Going back to the masses of vector bosons we note that in the limit $v_R\gg v$, where $v$ is any VEV ($v_L$, $k_1$, $k_2$, $k^{\prime}_1$, or $k^{\prime}_2$ ), we obtain from Eq.~(\ref{masszs2})
\begin{equation}
M^2_{Z_1}\approx \frac{g^2}{4\cos^2\theta}  \left(K+\frac{v^2_L}{2}\right),\quad M^2_{Z_2}\approx \frac{g^2+g^{\prime2}}{4}v^2_R,
\label{masszs3}
\end{equation}
and we see from (\ref{massws22}) and (\ref{masszs3}) that
\begin{equation}
M_{Z_1}\approx \frac{M_{W_1}}{\cos\theta}+\mathcal{O}(x). 
\label{ratio1}
\end{equation}
Notice that only in the limit $v_R\to\infty$ does the angle $\theta$ in this model have a relation with the $\theta_W$ of the SM. However,
it is important that $v_R$ is kept to be large but finite in order to obtain a lower bound on the right-handed vector bosons,
$W_2$ and $Z_2$~\cite{Bhattacharyya:1992iu} and the respective coupling with fermions. If $\chi_L$ is an inert doublet, we simply
put $v_L = 0$, or, equivalently, $y=0$, in the above expressions.

Using the exact results in Eqs.~(\ref{massws2}) and (\ref{masszs2}), using $r=g^\prime/g\approx0.6355$, $k_2\sim k^\prime_2\approx v_{SM}/\sqrt{2}$ ($z=x/2,y=0$ ) where $v_{SM}= 246$ GeV, and  $M_W/M_Z=0.88147\pm0.00013$~\cite{pdg2020}, we obtain (using 2$\sigma$ value of that ratio) that $v_R>24\,\textrm{TeV}$. With this lower limit on $v_R$, we can calculate the lower limit for the masses of $W^+_2$ and $Z_2$, using Eqs.~(\ref{massws2}) and (\ref{masszs2}), respectively, obtaining (in TeV)
\begin{equation}
M_{W_2}>7.2,\quad M_{Z_2}>9.28,
\label{masswz3}
\end{equation}
and the mixing angle $W_L-W_R$ defined in Eq.~(\ref{ws2}) has an upper limit $\sin\xi<10^{-4}$. Recent analysis comparing the
experimental limits to the theoretical calculations for the total $W_2$ resonant production and the decay $W_2\to WZ$ implies
that mixing angle $\xi$ is between $10^{-4}$ and $10^{-3}$, excluding $\xi\simeq 6\times 10^{-6}$~\cite{Serenkova:2019zav}.

\section{Yukawa interactions and fermion masses}
\label{sec:yukawa}

The Yukawa interactions in the lepton sector are given by
\begin{equation}
-\mathcal{L}_Y=\bar{L^\prime}_L(G\Phi_1+ F\tilde{\Phi}_1) L^\prime_R+\bar{L^\prime}_R(\Phi^\dagger_1G^\dagger+ \tilde{\Phi}^\dagger_1F^\dagger)L^\prime_L,
\label{yukawa1}
\end{equation}
where $L^\prime$ and $R^\prime$ are defined in Sec.~\ref{sec:model} and we have omitted generations indices. Since $\Phi_i\leftrightarrow \Phi^ \dagger_i$ under the left-right symmetry, then $G^\dagger=G$ and $F^\dagger=F$.

With these interactions and the vacuum alignment, the mass matrices in the lepton sector are
\begin{equation}
M^\nu=G\frac{k_1}{\sqrt2}+F\frac{k^{\prime*}_1}{\sqrt2},\quad M^l=G\frac{k^\prime_1}{\sqrt2}+F\frac{k^ *_1}{\sqrt2}.
\label{leptonmassmatrices}
\end{equation}
A similar expression arises in the quark sector but now $\Phi_1\to\Phi_2$ and $(\nu^\prime_{L,R},l^\prime_{L,R})
\to(u^\prime_{L,R},d^\prime_{L,R})$. We recall that the $\mathbb{Z}_2\otimes \mathbb{Z}^\prime_2$ symmetry forbids the coupling of the bidoublet $\Phi_2$ with leptons and $\Phi_1$ with quarks.

Primed fields denote symmetry eigenstates and unprimed ones, mass eigenstates.  In general $G,F$ and VEVs are complex, and the mass matrices are diagonalized by biunitary transformations as follows:
\begin{eqnarray}
V^{l\dagger}_LM^l V^l_R=\hat{M}^l,\quad U^{\nu \dagger}_L M^\nu U^ \nu_R=\hat{M}^\nu,
\label{vlunu}
\end{eqnarray}
where $\hat{M}^l=\textrm{diag}(m_e,m_\mu,m_\tau)$ and $\hat{M}^\nu=\textrm{diag}(m_1,m_2,m_3)$ for charged leptons and neutrinos respectively.

To give an appropriate mass to the quarks, we have to introduce the  bidoublet $\Phi_2$, and it is possible to implement the analysis as in Ref.~\cite{Senjanovic:2014pva}. Notice that this means that the neutral scalar with VEV and mass about 174 and 125 GeV, respectively, is part of this bidoublet. 

We will assume that $k^\prime_1=0$ (see Sec.~\ref{sec:potential}) and in this case the lepton mass matrices, from (\ref{leptonmassmatrices}), are given by
\begin{equation}
 M^\nu_{ab}=G_{ab}\frac{k_1}{\sqrt2}, \quad M^l_{ab}=F_{ab}\frac{k^*_1}{\sqrt2},
\label{leptonmassmatrices2} 
\end{equation}
where $G$ and $F$ are symmetric complex matrices that are diagonalized by the biunitary transformation in Eq.~(\ref{vlunu}). Hereafter, we will consider, just for the sake of simplicity, all VEVs being real.

From these matrices and the lepton measured masses, we found the Yukawa coupling matrices
\begin{equation}
G=\frac{\sqrt2}{k_1}U^ {\nu}_L\hat{M}^\nu U^{\nu\dagger}_R,\quad F=\frac{\sqrt2}{k^{*}_1}V^{l}_L \hat{M}^l V^{l\dagger}_R,
\label{yukawac}
\end{equation}
and we use for numerical calculations $\vert k_1\vert=2$ GeV, since this VEV is the only one for generating the lepton masses. 
For the sake of simplicity, we work in the basis in which the charged lepton mass matrix is diagonal and consider that the matrices $G$ and $F$ and all VEVs
are real. In this case  $U^\nu_L=U^\nu_R\equiv U^\nu$, and $U^\nu=V^L_{PMNS}=V^R_{PMNS}\equiv V_l$, and we have
\begin{equation}
G=\frac{\sqrt2}{k_1}V_l\hat{M}^\nu V^\dagger_l,\quad F=\frac{\sqrt2}{k_1}\hat{M}^l,
\label{app1}
\end{equation}
the unitary matrix $V_l$ being parametrized in the same way for Dirac particles. We use the PDG parametrization for Dirac neutrinos, for the interactions with $W^+_{L,R}$:
\begin{eqnarray}
V_l=\left(
\begin{array}{ccc}
c^l_{12}c^l_{13} & s^l_{12}c^l_{13} & s^l_{13}\\
-s^l_{12}c^l_{23}-c^l_{12}s^l_{13}s^l_{23}& c^l_{12}c^l_{23}-s^l_{12}s^l_{13}s^l_{23}& c^l_{13}s^l_{23}\\
s^l_{12}s^l_{23}-c^l_{12}s^l_{13}c^l_{23} & -c^l_{12}s^l_{23}-s^l_{12}s^l_{13}c^l_{23}  & c^l_{13}c^l_{23}
\end{array}
\right),
\label{pmns}
\end{eqnarray}
with $s^l_{ij}=\sin\theta^l_{ij},\cdots$ and where we have considered $\delta_l=0$.

In this case the Yukawa interactions are given by
\begin{eqnarray}
-\mathcal{L}^Y_l&=&\frac{\sqrt2}{k_1}\{\bar{\nu}_L[ (\hat{M}^\nu\phi^0_1+V^\dagger_l\hat{M}^lV_l\eta^{0*}_1)\nu_R
+(\hat{M}^\nu V^\dagger_l \eta^+_1-V^\dagger_l\hat{M}^l\phi^+_1)l_R] \nonumber \\ &+&
\bar{l}_L[(V_l\hat{M}^\nu \phi^-_1-\hat{M}^lV_l\eta^-_1)\nu_R+(V_l\hat{M}^\nu V^\dagger_l \eta^0_1+\hat{M}^l\phi^{0*}_1)l_R]\}\nonumber \\ &+&H.c.,
\label{yukawa6}
\end{eqnarray} 
with $V_l$ given in (\ref{pmns}). Notice that, in this case (in the basis in which charged leptons are diagonal), the Higgs $\phi^0_1$ is the one whose
couplings with charged leptons are proportional to their respective masses, and the couplings with $\eta^0_1$ are suppressed by the neutrino masses in
the charged lepton sector. In the neutrino sector, the situation reverses: The enhanced interactions are those with $\eta^0_1$ since they are proportional to the charged lepton masses. For instance, the vertex $\bar{\nu}_{3R}\nu_{1L}\eta^{0*}_1$ has the strength proportional to $s^l_{13}c^l_{13}c^l_{23}m_\tau$ and $\eta^0_1$ can decay throughout its mixing with the other neutral scalar, into two of the other particles, bosons or fermions. 

For completeness, we write the Yukawa interactions in the quark sector (with their mass matrices diagonalized by the unitary matrices $V^u_{L,R}$ and $V^d_{L,R}$ with $V^L_{CKM}=V^R_{CKM}=V^{u\dagger}_LV^d_L$):
\begin{eqnarray}
-\mathcal{L}^Y_q&=& \frac{\sqrt2}{k_1}\{  \bar{u}_LV^{u\dagger}[ (G_qV^u\phi^0_2+F_q\eta^{0*}_2)V^u\,u_R+(G_q\eta^+_2-F_q\phi^+_2)V^dd_R]\nonumber \\&+&
\bar{d}_LV^{d\dagger}[(G_q\phi^-_2-F_q\eta^-_2)V^u\, u_R+(G_q\eta^0_2+F_q\phi^{0*}_2)V^dd_R]\}+H.c.
\label{yukawa3}
\end{eqnarray} 
In the quark sector we shall not consider the solution $k^\prime_2=0$ since  for the case of generalized parity, $\mathcal{P}$, it has
been shown that $k^\prime_2\ll k_2$ is ruled out by the $C\!P$-violating parameters $\epsilon$ and $\epsilon^\prime$; however, 
this hierarchy is allowed in the case of generalized $\mathcal{C}$~\cite{Bertolini:2014sua}.  

Notice that there are flavor-changing neutral currents (FCNCs) mediated by scalars in both lepton and quark sectors.
However, the existence of these processes in the present model implies only that the constraints already obtained in the minimal
version of the model i.e.,  one bidoublet and two doublets, have to be reviewed. For instance, the current data and the
contributions related to the renormalization of the flavor-changing neutral Higgs tree-level amplitude which are needed in order
to obtain gauge-independent results was done in Ref.~\cite{Bertolini:2014sua}. In the minimal model there are four neutral scalars while in the present model there are six (with two bidoublets) or eight (three bidoublets) and it means that there are more amplitude mediated by neutral scalars at tree and one-loop level than those in Ref.~\cite{Bertolini:2014sua}. Doing this analysis is outside the scope of this paper. 

\section{Fermion-vector boson interactions}
\label{sec:leptonsgauge}

The covariant derivatives are given by
\begin{equation}
(\mathcal{D}_{\mu L_l(R_l)}) L^\prime_l (R^\prime_l)=\left(\partial_\mu+i\frac{g}{2}\vec{\tau}\cdot\vec{W}_{\mu L(R)}-i\frac{g^\prime}{2}B_\mu\right)L^\prime_l (R^\prime_l),
\label{dc1}
\end{equation} 
and similarly for quarks. The left lepton-gauge boson interaction is obtained from $\mathcal{L}~=~\bar{L}^\prime_l\gamma^\mu \mathcal{D}_{\mu L}L^\prime_l$ and similarly for the right-handed doublets.  

\subsection{Charged currents}
\label{subsec:cc}

The charged current interactions in the mass eigenstates basis are given by the Lagrangian 
\begin{eqnarray}
\mathcal{L}^l_W&=&-\frac{g}{2}\left[e^{i\phi_l}\overline{\nu_L}\gamma^\mu V_l l_L W^+_{L\mu}+\overline{\nu_R}\gamma^\mu V_l l_R W^+_{R\mu} \right]+H.c.\nonumber \\&=&
-\frac{g}{2}\left[(e^{i\phi_l}c_\xi J^{l\mu}_L +s_\xi J^{l\mu}_R ) W^+_{1\mu}+
(-e^{i\phi_l}s_\xi J^{l\mu}_L +c_\xi J^{l\mu}_R ) W^+_{2\mu} \right]+H.c.,\nonumber \\&&
\label{ci1}
\end{eqnarray}
and we have used Eq.~(\ref{ws1}); here $J^{l\mu}_L=\overline{\nu_L}\gamma^\mu V_l l_L$ and $J^{l\mu}_R=\overline{\nu_R} \gamma^\mu  V_l l_R$.  

In the quark sector,
\begin{eqnarray}
\mathcal{L}^q_W&=&-\frac{g}{2}\left[e^{i\phi_q}\bar{u}_L\gamma^\mu V_{CKM} l_L W^+_{L\mu}+\bar{u}_R\gamma^\mu V_{CKM} d_R W^+_{R\mu} \right]+H.c.\nonumber \\&=&
-\frac{g}{2}\left[(e^{i\phi_q}c_\xi J^{q\mu}_L +s_\xi J^{q\mu}_R ) W^+_{1\mu}+
(-e^{i\phi}s_\xi J^{q\mu}_L +c_\xi J^{q\mu}_R ) W^+_{2\mu} \right]+H.c.,\nonumber \\&&
\label{ci2}
\end{eqnarray}
with $J^{q\mu}_L=\bar{u}_L\gamma^\mu V_{CKM} d_L$ and $J^{q\mu}_R=\bar{u}_R \gamma^\mu V_{CKM}d_R$ with $V_{CKM}$ being the same as in the left-handed sector with three angles and one physical phase. 

In the general case where the Yukawa couplings in Eq.~(\ref{leptonmassmatrices}) are complex, the right-handed CKM matrix is different from the left-handed one. This case was considered in Ref.~\cite{Senjanovic:2015yea}. 

Here, the introduction of the phase $\phi_l$ and $\phi_q$ in Eqs.~(\ref{ci1}) and (\ref{ci2}), respectively, needs an explanation. In the mixing  matrix for $n$ Dirac fermions, $2n-1$ phases are absorbed in the Dirac fields, since one is a global phase. In the SM, this is enough, because there is only one charged current and the global phase never appears in amplitudes. However, in this sort of model there are also right-handed charged currents and there is a relative global phase between both charged currents. This phase is $\phi_l$ for leptons and $\phi_q$ for quarks.

\subsection{Electromagnetic interactions}
\label{subsec:emi}

The interaction with the photon arises from the projection of $W_{3L},W_{3R}$ and $B$ over $A$ using the matrix in Eq.~(\ref{nb1}). Then, it is possible to verify that the electric charge is written in terms of $g$ and $g^ \prime$ as 
\begin{equation}
e=\frac{gg^\prime}{\sqrt{g^2+2g^{\prime\,2}}},
\end{equation}
where we obtain
\begin{equation}
\frac{1}{e^2}=\frac{2}{g^2}+\frac{1}{g^{\prime2}},\quad\quad \frac{1}{g^2_Y}=\frac{1}{g^2}+\frac{1}{g^{\prime 2}},
\label{ratios}
\end{equation}
where $g_Y$ is the coupling constant of the SM.
These relations are valid only at the energy scale at which $g_L=g_R\equiv g$.
Hence, we have the results written in Eq.~(\ref{charge}).

From the igualities given in Eq.~(\ref{charge}), we have $r=g^\prime/g=\sin\theta/\sqrt{\cos\theta}$, obtaining also that
\begin{equation}
\frac{g^{\prime2}}{g^2}=\frac{s^2_\theta}{1-2s^2_\theta}.
\label{ratios2}
\end{equation}
The model has a Landau-like pole in $g^\prime$ when $s^2_\theta=1/2$ but it occurs at energies larger 
than the Planck scale. However, this implies only that the energy scale at which $g_L(\mu)=g_R(\mu)$ must be below the 
scale at which $s^2_\theta(\Lambda)=1/2$, $\mu<\Lambda$.

\subsection{Neutral currents}
\label{subsec:nc}

Next, we parametrize the neutral interactions of a fermion with the $Z_{1\mu}$ and $Z_{2\mu}$ neutral bosons as follows:
\begin{equation}
\mathcal{L}_{NC}=-\frac{g}{2\cos\theta}\sum_i\bar{\psi}_i\gamma^\mu[(g^i_V-g^i_A\gamma^5)Z_{1\mu}+(f^i_V-f^i_A\gamma^5)Z_{2\mu}]\psi_i,
\end{equation} 
Defining:
\begin{equation}
g^f_V=\frac{1}{2}(a^f_L+a^f_R), \quad g^f_A=\frac{1}{2}(a^f_R-a^f_L),
\label{defas}
\end{equation}
where $a^f_L$ and $a^f_R$ are the couplings of the left- and right-handed components, respectively, of a fermion $f$, and considering the case when the VEV of the doublet $\chi_L$ is not zero, $v_L\not=0$, using (\ref{nb1}),
$r=s_\theta /\sqrt{c_{2\theta} }$, and Eq.~(\ref{almost}), we obtain
\begin{eqnarray}
&& a^\nu_L\approx 1+\frac{t^2_\theta c_{2\theta}}{c_\theta}\left( x-\frac{s^2_\theta}{c_{2\theta}}y\right),\;\;
\hspace{0.7cm} a^\nu_R\approx \frac{c_{2\theta}}{c^2_\theta }\left( x-\frac{s^2_\theta}{c_{2\theta}}y\right),\nonumber \\ &&
a^l_L\approx c_{2\theta}\left[1-\frac{t^2_\theta}{c^2_{\theta}}\left(x-\frac{s^2_\theta}{c_{2\theta}}y \right)\right], \quad %\nonumber \\ && 
a^l_R\approx-2s^2_\theta+\frac{x}{c^4_\theta}(s^4_\theta+c^2_{2\theta})-yt^2_\theta\frac{c_{2\theta}}{c^2_\theta}.
\label{gvga1}
\end{eqnarray}
In eq. (\ref{defas}) we get
\begin{eqnarray}
&& g^{\nu_l}_V\,\approx\,\frac{1}{2}+\frac{c_{2\theta}}{2\,c^4_{\theta}}\left(x-y\,\frac{s^2_{\theta}}{c_{2\theta}}\right),\,\quad \quad \quad \quad \quad \quad \quad 
g^{\nu_{\ell}}_A\,\approx\,\frac{1}{2}-\frac{c^2_{2\theta}}{2\,c^4_{\theta}}\left(x-y\,\frac{s^2_{\theta}}{c_{2\theta}}\right),  \nonumber\\&&
g^l_V\,\approx\,\frac{1}{2}(-1+4s^2_\theta)\left\{1+\frac{c_{2\theta}}{c^4_{\theta}}\left(x-y\frac{s^2_{\theta}}{c_{2\theta}}\right)\right\},\quad
g^l_A\,\approx\,-\frac{1}{2}+\frac{c^2_{2\theta}}{2\,c^4_{\theta}}\,\left(x-y\frac{s^2_{\theta}}{c_{2\theta}}\right).
\label{gvgaleptons}
\end{eqnarray}
	
Notice that, when $v_R\to\infty\,\,(\mbox{that is}\,\, x,y\to 0)$, we obtain
\begin{eqnarray}
g^\nu_V=g^\nu_A=\frac{1}{2},\quad g^l_V=-\frac{1}{2}+2s^2_\theta,\,\,\,\,g^l_A=\frac{1}{2},
\end{eqnarray}
and the same happens with the coefficients of the quark sector in this limit, obtaining
\begin{eqnarray}
&& g^u_V=\frac{1}{2}-\frac{4}{3}s^2_\theta,\quad g^u_A=\frac{1}{2},\qquad\quad g^d_V=-\frac{1}{2}+\frac{2}{3}s^2_\theta,\quad g^d_A=-\frac{1}{2},\nonumber \\&&
f^u_V=\frac{1}{2}-\frac{4}{3}s^2_\theta,\quad f^u_A=-\frac{1}{2}c_{2\theta},\quad f^d_V=-\frac{1}{2}+\frac{2}{3}s^2_\theta,\quad f^d_A=\frac{1}{2}c_{2\theta}.
\end{eqnarray}
It is worth noting that in the quark sector the vector couplings are the same for $Z_1$ and $Z_2$.
We see once again that only when $v_R$ is strictly infinite can we identify, at tree level, the angle $\theta$ with $\theta_W$ of the SM.
Assuming the measured values $g^l_V=0.03783\pm0.00041$ does not imply a stronger lower bound on $v_R$ and the $W_2$ and $Z_2$ masses, which was obtained from the $M_W/M_Z$ ratio in Eq.~(\ref{masswz3}).  

Recently, the CMS Collaboration using $W_2\to B+t$ or $W_2\to T+b$ [$T$ and $B$ are vectorlike quarks (VLQs)] excluded a $W_2$ with a mass below
1.6 TeV at $95\% $ C.L. assuming equal branching ratios for the $W^\prime$ boson to $tB$ and $bT$ and 50\% for each VLQ 
to $qH$, where $H$ is a neutral scalar~\cite{Sirunyan:2018fki}. If $T$ and $B$ are the known $t$ and $b$ quarks and assuming $W_2$ with coupling 
to the SM particles equal to the SM weak coupling constant, masses below 3.15 TeV are excluded at the 95\% confidence
level~\cite{Aaboud:2018jux}. 

Furthermore, if right-handed gauge bosons decay
into a high-momentum heavy neutrino and a
charged lepton, the LHC has excluded values of the $W_2\sim W_R$ smaller than 3.8–5 TeV for $N_R$ in the mass range 
0.1–1.8 TeV~\cite{Aaboud:2019wfg}. Of course, if there are no extra quarks like $ T $ and $ B $ and neither heavy right-handed
neutrinos, these restrictions for the mass of $ W_R $ are not valid anymore.

Only for illustration, we give the partial width at tree level, neglecting the fermion masses, and with $M_{W_2}=7.8$ TeV 
\begin{equation}
\Gamma(W^+_2\to l^+\nu) \approx \frac{G_FM^2_{W_1}M_{W_2}}{6\pi \sqrt{2}}\sim \;31.57\;\textrm{GeV}.
\label{wr}
\end{equation}
Adding over all fermions, it means a full width $\Gamma\sim 94{.71}$ GeV. Compare this with the case of the $W$ of the SM, $\Gamma_W=2.085\pm0.042$ GeV~\cite{pdg2020} where $l$ denotes any of the charged leptons i.e., $l=e,\mu,\tau$ without a sum over them. 

For the $Z_2$ and also neglecting the fermion masses we have 
\begin{equation}
\Gamma(Z_2\to f\bar{f})\simeq N_c\left [(a^f_L)^2+ (a^f_R)^2\right]\frac{G_FM^2_WM_{Z_2}}{24\pi},
\label{zp1}
\end{equation}
for leptons $N_c=1$ and for quarks $N_c=3$. In the case of leptonic decay, using the couplings in Eq.~(\ref{gvga1}),  
we have~$\Gamma(Z_2\to~l^-l^+)\sim 3.79$ GeV for any of the three charged leptons, if $M_{Z_2}=9.28$ TeV, with $\Gamma(Z\to 
l^-l^+)~=~83.984\pm0.086$ MeV~\cite{pdg2020}. %\cite{ATLAS:2014wra}

Notice that scalar doublets $\chi_{L,R}$ do not couple to fermions and we will assume that vacuum alignment is such that $v_L=0$;
therefore, this scalar field does not contribute to the gauge boson masses, and, hence, $\chi_L$ is an inert doublet~\cite{Kalinowski:2019cxe}. In 
this case, the inert character is protected by the left-right symmetry.

\section{Lepton masses and mixing}
\label{sec:lmasses}

Here we will obtain, assuming the measured matrix elements of the PMNS matrix, the Yukawa couplings to generate the correct
charged lepton and neutrino masses.  First, we consider the present case, i.e., two bidoublets, then we briefly discuss 
the case with three bidoublets.  

\subsection{The two-bidoublet case}
\label{subsec:2bd}

First, we neglect $C\!P$ violation, which means that the matrices $F$ and $G$ are real  and we consider $F$ diagonal, with $U^\nu=V_l$; see Eq.~(\ref{pmns}). 
Concerning the lepton masses, in the charged lepton sector we will use the central values given in PDG~\cite{pdg2020},  and 
in the neutrino sector we will use the several possibilities:

\noindent
(i) Normal mass hierarchy (NH), $m_1\ll m_2<m_3$.
\begin{equation}
m_1=0,\quad m_2\simeq (\Delta m^2_{21})^{1/2}~\simeq~0.0086 \;\textrm{eV},\quad m_3\simeq\vert\Delta m^2_{31}\vert^{1/2}\simeq0.0506 \;\textrm{eV}.
\label{nh}
\end{equation}
(ii) The inverted hierarchy (IH) $m_3\ll m_1<m_2$.
\begin{equation}m_3=0,\quad 
m_1\simeq 0.0497, \quad m_2~\simeq~0.0504 \;\textrm{eV}, 
\label{ih}
\end{equation}
(iii) The quasidegenerate case (QD) $m_1\simeq m_2\simeq m_3\simeq m_0$~\cite{pdg2020}.
\begin{equation}
m_j\gg  \vert \Delta m^2_{(31)32}\vert^{1/2},\quad m_0\lesssim 0.10  \;\textrm{eV}.
\label{degenerado}
\end{equation}
Recall that in the case we are considering here, the PMNS mixing matrix is given by $V_l$.

Using the numerical values for the neutrinos masses in  Eq.~(\ref{nh}), the PDG's angles,
\begin{eqnarray}
s^2_{12}=0.307,\quad 
s^2_{23}=0.0.512 \;(\textrm{normal order, octant I)},\quad 
s^2_{13}=0.00218,\;
k_1=2\,\mbox{GeV},
\label{pdg}
\end{eqnarray}
\noindent
(i) From Eq.~(\ref{app1}),  in the normal hierarchy 
$\hat{M}^\nu=\textrm{Diag}(0,\sqrt{\Delta m^2_{12}}, \sqrt{\Delta m^2_{31}}) $, and using the matrix given in Eq.~(\ref{pmns}), we obtain (up to a factor $10^{-11}$) 
\begin{eqnarray}
&& G_{11}\approx0.2606,\,
G_{12}\approx0.5481\, \,
G_{13}\approx0.1474,\,
G_{21}\approx0.5481,\nonumber \\&&
G_{22}\approx1.9583,\,
G_{23}\approx1.5418,\,
G_{31}\approx0.1474,\,
G_{32}\approx1.5418,\nonumber \\&&
G_{33}\approx1.8671.
\label{gsnh}
\end{eqnarray}
\noindent
Similarly, in the charged lepton sector we have~\cite{pdg2020}
\begin{equation}
F_e\approx 3.613\times10^{-4},\quad F_\mu\approx 0.075,\quad  F_\tau\approx1.25.
\label{fs}
\end{equation}
(ii) Using the inverse mass hierarchy in Eq.~(\ref{ih}), we obtain  (up to a factor $10^{-11}$):
\begin{eqnarray}
&&G_{11}\approx3.4526,\,
G_{12}\approx-0.3530,\,
G_{13}\approx-0.3762,\,
G_{21}\approx-0.3530,\nonumber \\ &&
G_{22}\approx1.7677,\,
G_{23}\approx-1.7352,\,
G_{31}\approx-0.3762,\,
G_{32}\approx-1.7352,\nonumber \\ &&
G_{33}\approx1.8568.
\label{gsih}
\end{eqnarray}
(iii) In the quasidegenerate case, in (\ref{degenerado}),  we obtain (up to a factor $10^{-11}$),
\begin{eqnarray}
G_{11}=G_{22}=G_{33}\approx7{.}0711,
\label{gsquasi}
\end{eqnarray}
in this case, all the other $G$'s vanish for all practical purposes. 

Although this model with two bidoublets contains a fine adjustment as in the SM, which is avoided if we introduce a third bidoublet, 
this would be the price to pay for having Dirac neutrinos and only the known charged leptons plus the right-handed neutrinos.
However, we will show that when a third bidoublet is considered, it seems possible to avoid a fine-tuning in the lepton masses.

\subsection{Three-bidoublets case}
\label{subsec:3bd}

It is interesting that one of the natural hierarchy in field theories are those in which the VEVs are responsible
by the spontaneously breaking of symmetries. This is because their values depend on the vacuum alignment and heavy scalars may have 
small VEVs. Probably this was first noted by Ma~\cite{Ma:2000cc} and we have seen an example in Sec.~\ref{sec:potential} in the case 
of $k^\prime_1$. Moreover, as we have emphasized before, we already do not know the number and sort of scalars and we can think 
of an extension of the present model in which three bi-doublets (and the two doublets $\chi_{L,R}$) are introduced.

In this case, the sector of the model which is more affected by the existence of a third bi-
doublet is the Yukawa one. Let us denote $\Phi_\nu,\Phi_l$ and $\Phi_q$ the three bi-doublets.  We denote the respective VEVs
$\sqrt{2}\langle \Phi_\nu\rangle=\textrm{Diag}(k_\nu\;k^\prime_\nu)$,  $\sqrt{2}\langle \Phi_l\rangle=\textrm{Diag}(k_l\;k^\prime_l)$, 
and $\sqrt{2}\langle \Phi_q\rangle=\textrm{Diag}(k_q\;k^\prime_q)$.

We introduce the discrete symmetry $D$ under which~\cite{Branco:1978bz}
\begin{eqnarray}
D:\quad \Phi_\nu,\Phi_l\to  -i\Phi_\nu,-i\Phi_l,\quad R^\prime_l\to iR^\prime_l,
\label{3bidoublets}
\end{eqnarray}
and all the other fields stay invariant under $D$. In this case, the Yukawa interactions are written as
\begin{equation}
\mathcal{L}_Y=\bar{L}^\prime_l (G^\nu  \Phi_\nu+ G^l  \Phi_l) R^\prime_l+\bar{Q}^\prime_L(G^q \Phi_q+F^q\tilde{\Phi}_q)Q^\prime_R+H.c.
\label{newyuka}
\end{equation}
Notice that the $D$ symmetry forbids the interactions like $\bar{L}^\prime_l\tilde{\Phi}_\nu R^\prime_l$ and
$\bar{L}^\prime_l\tilde{\Phi}_l R^\prime_l$, where $\tilde{\Phi}=\tau_2\Phi^*\tau_2$.
Although it is out of the scope of this work to analyze the scalar potential and its spectra we note that it may be possible to 
have a vacuum alignment in which VEVs are hierarchically: $k_\nu, k^\prime_\nu,k_l\ll k^\prime_l\ll k_q,k^\prime_q$, then neutrino
masses arise from $k_\nu$, and the charged lepton masses from $k^\prime_l$ (these leptons receive a small contributions from $k^\prime_\nu$).
In this situation the mass matrices are given by
\begin{equation}
M^\nu\approx G^\nu\frac{k_\nu}{\sqrt2},\quad M^l\approx G^l\frac{k^\prime_l}{\sqrt2},\quad M^u=G^q\frac{k_q}{\sqrt2}+F^q\frac{k^{\prime*}_q}{\sqrt2},\quad M^d=G^q\frac{k^\prime_q}{\sqrt2}+F^q\frac{k^{*}_q}{\sqrt2},
\end{equation}
where the quark mass matrices are the same as in most LR symmetric models.
If $k_\nu\stackrel{>}{\sim}\sqrt{ \Delta m^2_{31}}$ then all entries of the matrix $G^\nu$ are of order of unity. The same happens in the Yukawa couplings in the charged lepton sector if $k^\prime_l \stackrel{>}{\sim}\,m_\tau $.
Recall that hierarchy among VEVs are more easily justified than in the Yukawa couplings. The quark sector follows as usual. 
An interesting possibility is when an $S_3$ symmetry is introduced.
This has been done in the quark sector by Das and Pal~\cite{Das:2018rdf}, and it is possible to do the same in the lepton sector.

We illustrate how hierarchy between the VEVs can arise considering the SM with two scalar doublets with $Y=+1$,  $H_i,\;i=1,2$ with $\langle H^0_1\rangle=v$ and  $\langle H^0_2\rangle=u$. Introducing the quadratic non-Hermitian term in the scalar potential $\mu^2_{12}(H^\dagger _1 H_2+H^\dagger_2 H_1)$ the constraints equations have the form \cite{Ma:2000cc}
\begin{equation}
v[\mu^2_1+\lambda_1 v^2+(\lambda_3+\lambda_4)u^2]+\mu^2_{12}u=0 , \quad u[\mu^2_2+\lambda_2u^2+(\lambda_3+\lambda_4)v^2 ]+\mu^2_{12}v=0,
\label{cef}
\end{equation}
where $\lambda_{1,2,3,4}$ are quartic coupling constants. If $\mu^2_1<0$, $\mu^2_2>0$, and $\vert \mu^2_{12}\vert \ll \mu^2_2$, we have
\begin{equation}
v^2\simeq -\frac{\mu^2_1}{\lambda_1},\quad u\simeq -\frac{\mu^2_{12} v}{\mu^2_2+(\lambda_3+\lambda_4)v^2}.
\label{obam}
\end{equation}
We see that $u\ll v$ is possible. A similar mechanism may be at work in the present model but the
full scalar potential with three bidoublets and two doublets is rather complicated and needs a separately study. In fact, these is what we have done in Sec.~\ref{sec:potential} for the case of two bidoublet if we impose the $\mathbb{Z}_5$ symmetry.

\section{Some phenomenological consequences}
\label{sec:feno}

Many of the features of the present model are as those in multi-Higgs models. For instance, FCNC mediated by scalars, several $C\!P$-violating phases, etc. The existence of FCNCs in the scalar sector has several phenomenological consequences, among others, it means that there are contributions to the muon anomaly $\Delta a_\mu$. 
For instance, taking the present data for the case of the muon  $a_\mu=(g_\mu-2)/2$ anomaly:
$\Delta a_\mu ~=~a^{exp}_\mu-a^{SM}_\mu=288(63)(49)\times10^{-11}$ which is about 3.7$\sigma$ below the experimental value~\cite{pdg2020}. In the present model there
are several possible contributions to $a_\mu$. Just for illustration consider the  contribution of a scalar or a 
pseudoscalar~\cite{Jegerlehner:2009ry}  
\begin{equation}
\Delta a^ \mu_X(f)= \frac{m^2_\mu} {8\pi^2m^2_X } \vert \mathcal{O}\vert^ 2\int_0^ 1dx\,\frac{ Q_X(x)}{(1-x)(1-\lambda^2_X)+(\epsilon_f\lambda_X)^2x},
\label{mdm}
\end{equation}
with $X=S,A$, $\epsilon_f=m_f/m_\mu$, where $f$ is the fermion in the internal line, $\lambda_X=m_\mu/M_X$ (if $f$ is lepton tau)
and $Q_S(x)=x^2(1+\epsilon-x)$ for a scalar $S$, and $Q_A=x^2(1-\epsilon-x)$ for a pseudoscalar $A$; $\mathcal{O}$ denotes
a matrix element in the scalar or pseudoscalar sector and we use, for simplicity, $\vert \mathcal{O}\vert =1$.  In order to fit the muon and 
electron $g-2$ anomalies~\cite{parker2018} we need $m_S\stackrel{>}{\sim}4.318$ TeV 
and $m_A\stackrel{>}{\sim}4.321$ TeV~\cite{DeConto:2016ith}. However, lower masses are allowed if we consider the contributions
of all scalar and pseudoscalars in the model. We recall that vector boson contributions $W_1$ and $W_2$ are suppressed by neutrino masses
and the unitarity of the PMNS mixing matrices, where neutral vector bosons have diagonal interactions with leptons.

Below, we will consider mainly the difference with the case of the model with Majorana
neutrinos (with triplets in the scalar sector), in particular, when heavy neutrinos do exist,
with the present model with Dirac neutrinos.

\begin{itemize} 
\item[(i)] In the present case, there are no heavy neutrinos that can decay into a Higgs boson plus an active neutrino, $\nu_R\to H+\nu^\prime_L$. These processes are kinematically forbidden since neutrinos are the lightest particles in the model.

\item[(ii)]  Flavour-changing lepton number processes as $\mu\to e +\gamma$ and $\mu-e$ are suppressed because of the small neutrino masses. 
The case $\mu\to ee\bar{e}$ is discussed below. For instance, $\mu\to e+\gamma$ may occur through charged scalar or a $W^+$ where the branching ratio would 
(up to numerical factors $\sim\alpha$) be proportional to
\begin{equation}
 \left|\sum_{k=1}^3(V^*_l)_{ik}(V_l)_{jk}\frac{m^2_{\nu_{k}}}{M^2_W}\right|^2,
\end{equation}
where $V_l$ is the PMNS matrix in Eq.~(\ref{pmns}) and we have used the values given in Eq.~(\ref{pdg}); $M_W$ denotes the mass of $W_L$ or $W_R$. Plugging in numbers, we obtain branching ratios smaller than $10^{-48}$. The suppression is due the small neutrino masses and the GIM suppression factor (unitarity of the matrix $V$). 
Also, in the present model  there is not the (logarithmic) enhancement produced by the  doubly charge scalar bosons~\cite{Raidal:1997hq,Tello:2012qda}.  
\item[(iii)]  In the present model with Dirac neutrinos, there is no  neutrinoless double beta decay 
$(\beta\beta)_{0\nu}$ and other $\vert \Delta L\vert=2$  processes; the muon decay proccess $\mu\to ee\bar{e}$  are produced at tree level 
only by neutral scalars, as can be seen from the Yukawa interactions in Eq.~(\ref{yukawa6}). Among the flavor violation charged lepton decays,
this is the one which imposes the strongest restriction on the new physics scenarios. Recall that the present and future sensitivities in 
this decay are $10^{-12}$ and $10^{-16}$, respectively ~\cite{Galli:2019xop}. However, this process is proportional to (up to kinematic
factors) $\vert V^\dagger _LGV_R\vert^4/m^4_X$, where $V_L$ and $V_R$ are unitary matrices and $G$ are the Yukawa couplings in Eq.~(\ref{yukawa6}).
In the case of two bidoublets, the $G$ entries are as those in Eqs.~(\ref{gsnh}), (\ref{gsih}) and (\ref{gsquasi}), and all of 
them are rather small $\sim 10^{-11}$; hence their contributions are negligible. This is not the case if we consider three bidoublets where
there is no fine-tuning in the Yukawa couplings. In this case, the $\mu\to ee\bar{e}$ decay will impose direct constraints on these couplings. However, in this 
case Yukawa couplings of the order of $10^{-3}$ will suppress the factor $\vert V_L^\dagger GV_R\vert^4<10^{-12}$ and the decay 
$\mu\to ee\bar{e}$ can have a rate near to the experimental limit.
The latter case deserves a more detailed study.

\item[(iv)]  Keung-Senjanovic (KS) noted that the LHC offers an exciting possibility of seeing directly both LR symmetry restoration and lepton number violation production of same sign in charged lepton pairs plus jets: $pp\to W^+_R\to l^+_RN^c_L\to l^+_R W^-_Rl^+_R\to  l^+_Rl^+_R jj$ where  $(j)$ denotes jets~\cite{Keung:1983uu}. 

\item[(v)]   Instead of jets, we have another charged leptons and one neutrino;  the respective trileptons final state has been considered in Ref.~\cite{Helo:2018rll}.  However, if neutrinos are pure Dirac particles, as in the present model, there are no heavy right-handed neutrinos, and, hence, these sort of processes, since one of the vector bosons is $W_R$ and the other $W_L$; it needs a mass insertion in the neutrino internal line, and, hence the amplitude is proportional to the (active) neutrino mass and, for this reason, negligible. Of course, we can introduce triplets in order to have a seesaw type I or II mechanism, but we think that it is still interesting to study pure Dirac neutrinos in LR symmetric models. 

Trilepton processes occur in both Majorana and Dirac neutrino cases, and there are subprocesses like the following
\begin{eqnarray}
&& qq^\prime\to W^+_R\to \nu_Rl^+_{1L}\to W^+_R l^+_{1L} l^-_{2L} \to l^+_{1L}l^-_{2L}l^+_{3L}N_R, \;\;\;\;\;(a)\;\;\; \textrm{M}
\nonumber \\&&
qq^\prime\to W^+_L\to \nu_Ll^+_{1R}\to W^+_L l^+_{1R}l^-_{2L}\to l^+_{1R}l^-_{2L}l^+_{3R}(N_R)^c,\;\;(b)\;\;\; \textrm{M}\nonumber \\ &&
qq^\prime\to W^+_R\to \nu_R l^+_{1L}\to W^+_L l^+_{1L} l^-_{2L}\to l^+_{1L}l^-_{2L}l^+_{3R}\nu_L,\;\;\;\;\;\;\;\;(c)\;\;\; \textrm{D}\nonumber \\ &&
qq^\prime\to W^+_L\to \nu_L l^+_{1R}\to W^+_R l^+_{1R} l^-_{2R}\to l^+_{1R}l^-_{2R}l^+_{3L}\nu_R\;\;\;\;\;\;\;\;\;(d)\;\;\; \textrm{D}\nonumber \\ &&
qq^\prime\to W^+_R\to \nu_R l^+_{1L}\to W^+_R l^+_{1L}l^-_{2R}\to l^+_{1L}l^-_{2R}l^+_{3L}\nu_R(N_R),\;\,\,(e)\;\;\; \textrm{D,M}\nonumber \\ &&
qq^\prime\to W^+_L\to \nu_Ll^+_{1R}\to W^+_L l^+_{1R} l^-_{2L} \to l^+_{1R}l^-_{2L}l^+_{3R}\nu_L,\;\;\;\;\;\;\;\;(f)\;\;\; \textrm{D,M}.
\label{ksp}
\end{eqnarray}  
Above we denote $N_R$ a heavy right-handed neutrino and $\nu_R$ the right-handed component of a Dirac neutrino. Notice that the
processes (c) and (d) need a mass insertion or a Yukawa coupling and they are suppressed if neutrinos are pure Dirac. Hence, at least in 
principle, it is possible to use these processes to distinguish the Dirac from the Majorana case. When processes can occur in 
both Majorana and Dirac cases, the Dirac case is suppressed by the small neutrino masses. Of course, the Majorana case allows $\Delta L=2$, and the
Dirac case does not.

\item[(vi)]  Finally, but not least, we note that in LR or other models with a second charged current, there is a contribution to the electric dipole
moment (EDM) of an elementary particle, charged leptons or quarks (neutron), at the one-loop level. The new phase in the interactions with
the $W^\pm$ do not contribute at this level since in this diagram 
$C\!P$-violating phase cancels out because one vertex is the 
complex conjugate of the other, the diagram is proportional to $V_{CKM}V^*_{CKM}$ in the quark sector, and
the diagram is real~\cite{Ellis:1976fn}. However, if there is a second charged current as in LR symmetric models,
the relative phase $\phi_l$ and $\phi_q$ in Eqs.~(\ref{ci1}) and (\ref{ci2}) cannot be observed. In particular, it implies a contribution 
to the EDM of the quark:
\begin{equation}
\mu^q_E=e\frac{m_q}{M^2_{W_i}}c_\xi s_\xi \sin\phi_q\times\textrm{logarithmic corrections},
\end{equation}
where $M^2_{W_i}=M^2_{W_1,W_2}$, and the larger contribution is that of $W^+_1$. As an illustration, the EDM of a light quarks with $m_q\sim10^{-3}$ GeV and $M_{W_1}\sim80$ GeV, $s_\xi\sim10^{-3}$ we obtain $\mu^q_E\approx 3.08\times 10^{-26}\,\sin\phi_q\,\;e\,\textrm{cm}$, which is almost the present limit for the EDM of the neutron $\mu^n_E<0.30\times10^{-25}\,e\,\textrm{cm}$ C.L. 90\%. Hence, the phase $\phi_q$ is not restricted  with the present experimental data.  In the lepton sector  the most restricted EDM is that of the electron $\mu^e_E<0.87\times 10^{-28} e\,\textrm{cm}$, CL=10\%~\cite{pdg2020} which implies $\sin\phi_l<10^{-2}$. In the lepton sector the one-loop contribution (induced by the $W_L-W_R$  mixing) to the electron EDM is suppressed by the neutrino masses and the phase $\phi_l$ is not  suppressed by this observable. On the other hand, neutrinos have magnetic and electric dipole moments (induced by the mixing) which are not suppressed, since they are proportional to the heavy charged lepton. 

\end{itemize} 

\section{Breaking parity first}
\label{sec:parityfirst}
Any model beyond the Standard Model must match with that model at a given energy,
say, the $Z$ pole. In the SM coupling constants $g$ and $g_Y$ have different running with the energy.  In the case of LR symmetric models, the same happens with $g$ and $g^\prime\equiv g_{B-L}$ as was noted in Ref.~\cite{Chang:1984uy}. It means that we cannot keep $g_L=g_R$ for all energies, since quantum corrections imply a finite $\Delta g=g_L-g_R\not=0$. This is due to the fact that both constants feel different degrees of freedom. Hence, it is interesting to search for models with gauge symmetries as in Eq.~(\ref{simmetry}) but in which parity is broken spontaneously by nonzero VEVs~\cite{Chang:1984uy} or softly if quadratic terms in the scalar potential are different $\mu^2_L\not=\mu^2_R$ as in Ref~\cite{Mohapatra:1974gc}. 

Let us consider as in Ref.~\cite{Chang:1984uy}  the possibility that the symmetries in Eq.~(\ref{simmetry}) are broken spontaneously but in the following way: First, the parity $\mathcal{P}$ is broken at an energy scale $\mu_P$ by introducing a neutral pseudoscalar singlet, $\eta\sim(\mathbf{1},\mathbf{1},0)$ with $\eta\to -\eta$ under parity and $\langle\eta\rangle=v_\eta$. Then, the doublet $\chi_R$ breaks the $SU(2)_L\otimes SU(2)_R\otimes U(1)_{B-L}$ symmetry to $SU(2)_L\otimes U(1)_Y$. The relevant terms in the scalar potential involving the doublets $\chi_L,\chi_R$ and the isosinglet $\eta$ are the following:

\begin{equation}
\mu^2_\eta\eta^2+\lambda_\eta\eta^4+\mu^2_{LR}(\chi^\dagger_L\chi_L+\chi^\dagger_R\chi_R)+f\eta(\chi^\dagger_L\chi_L-\chi^\dagger_R\chi_R)+\lambda^\prime_\eta\eta^2(\chi^\dagger_L\chi_L+\chi^\dagger\chi_R)\subset V.
\label{eta1}
\end{equation}
At the energy $\mu_{\mathcal{P}}$, $\mu^2_\eta<0$ with $\langle\eta\rangle=v_\eta\simeq\mu_{\mathcal{P}}$, and all the other VEVs are still zero. We obtain
\begin{equation}
\mu^2_L=\mu^2_{LR}+fv_\eta+\lambda^\prime_\eta v^2_\eta,\quad \mu^2_R=\mu^2_{LR}-fv_\eta+\lambda^\prime_\eta v^2_\eta,
\label{eta2}
\end{equation}
with the singlet VEV $v_\eta=\sqrt{-\mu^2_\eta/2\lambda_\eta}$. Next, if $\mu^2_R<0$ and $\vert\mu^2_R\vert\ll v_\eta$, we have that $\langle\chi_R\rangle=v_R\not=0$. This leads to the interesting case in which the $SU(2)_R$ symmetry-breaking scale is induced by the 
parity-breaking scale as noted in Refs.~\cite{Chang:1984uy}. 
It happens also that $g_L\not= g_R$, for energies in the range $v_R<\mu<v_\eta$, and also $V^L_{PMNS}\not= V^R_{PMNS}$, with $V^L_{PMNS}=V^{l\dagger}_LU^\nu_L$,  $V^R_{PMNS}=V^{l\dagger}_RU^\nu_R$.  In this case we have to consider the most scalar potential involving two or three bidoublets, $\Phi_i$, two doublets $\chi_{L,R}$, and the singlet $\eta$. 

\section{Conclusions}
\label{sec:con}

In the context of the SM with three right-handed neutrinos the Yukawa couplings have the hierarchy (using the normal hierarchy) $\Delta y_{31}=(\Delta m^2_{31})^{1/2}/v_{SM}\approx2\times10^{-13}$ and 
$\Delta y_{21}=(\Delta m^2_{21})^{1/2}/v_{SM}\approx2\times10^{-14}$, which we compare with those couplings that in the present model are given in Eq.~(\ref{gsnh}). Although the latter values are smaller than the Yukawa sector in the charged lepton sector, see Eq.~(\ref{fs}), we note that the dispersion in the neutrino Yukawa couplings is in the range $0.25-2.2$ (up to a factor of $10^{-11}$). In this model, as in the old left-right symmetric models without scalar triplets and also no extra charged leptons, neutrinos gain arbitrary small masses. Notice, however that the Yukawa couplings are all almost of the same order of magnitude and about 2 order of magnitude larger compared with those  in the context of the SM with three right-handed neutrinos. However, we show in Sec.~\ref{sec:lmasses} how this fine-tuning in the lepton sector can be avoided at the price of introducing a third bidoublet and the discrete $D$ symmetry. In the latter case, all Yukawa couplings in the lepton and quark sector may be of the  order of $\mathcal{O}(1)$ if there is a hierarchy in the VEVs of the three bidoublets.

We stress that, since the early 1980s, most phenomenology of the left-right symmetric models includes triplets and Majorana neutrinos~\cite{Mohapatra:1979ia}. Since then, the model with the following scalar multiplets: one bi-doublet and two triplets was considered the minimal left-right symmetric model. There is no doubt that this proposal was, and still is, well motivated~\cite{Senjanovic:2016bya}. However, if the neutrinos ultimately turn out to be Dirac particles, all that effort will have been in vain.  For this reason we have to pay attention to Dirac neutrinos, even in the context of the left-right symmetric models. 

\section*{ACKNOWLEDGMENTS}

H.D. thanks CONCYTEC for financial support  and the IFT-UNESP for the kind hospitality where part of this work was done.
V.P. would like to thanks for partial financial support CNPq and FAPESP under the funding Grant No. 2014/19164-6 and last, but not least, the Faculty of Sciences of the Universidad Nacional de Ingenierıía (UNI) for the kind hospitality.

\end{document}